# Far- and Mid-infrared Emission and Reflectivity of Orthorhombic and Cubic ErMnO$_3$: polarons and bipolarons

Néstor E. Massa*,[1] Leire del Campo,[2] Karsten Holldack,[3] Vinh Ta Phuoc,[4] Patrick Echegut,[2] Paula Kayser,[5] and José Antonio Alonso[6]

[1] Centro CEQUINOR, Consejo Nacional de Investigaciones Científicas y Técnicas, Universidad Nacional de La Plata, Bv. 120 1465, B1904 La Plata, Argentina.

[2] CNRS-CEMHTI UPR3079, Univ. Orléans, F-45071 Orléans, France.

[3] Helmholtz-Zentrum für Materialien und Energie GmbH, Institut für Methoden und Instrumentierung der Forschung mit Synchrotronstrahlung (BESSYII) D-12489 Berlin, Germany.

[4] Groupement de Recherche Matériaux Microélectronique Acoustique Nanotechnologies-Université François Rabelais Tours, Faculté des Sciences & Techniques, F- 37200 Tours, France.

[5] Centre for Science at Extreme Conditions and School of Chemistry, University of Edinburgh, Kings Buildings, Mayfield Road, EH9 3FD Edinburgh, United Kingdom.

[6] Instituto de Ciencia de Materiales de Madrid, CSIC, Cantoblanco, E-28049 Madrid, Spain.

•e-mail: neemmassa@gmail.com

# Abstract

We report on the high-temperature evolution of far- and mid- infrared reflectivity and emissivity spectra of ambient orthorhombic $ErMnO_3$ from 12 K to sample decomposition above 1800 K. At low temperatures the number of phonons agrees with the predictions for orthorhombic space group $D_{2h}^{16}$-Pbnm (Z=4) and coexists with a paramagnon spin resonance and rare earth crystal field transitions. Increasing the temperature, a number of vibrational bands undergo profile broadening and softening approaching the orbital disordered phase where the orthorhombic O´ lower temperature cooperative phase coexists with cubic-orthorhombic O. O-$ErMnO_3$ undergoes a first order order-disorder transition into the perovskite cubic phase at $T_{cubic}$ ~1329 K ± 20 K where the three triple degenerate phonons allowed by the space group Pm-3m(Z=1) are identified.

At about 800 K, a quantitative small polaron analysis of the orthorhombic mid-infrared real part optical conductivity shows that antisymmetric and symmetric breathing modes sustain the strongest electron-phonon interactions. Above $T_{cubic}$ the bipolaron fingerprint profile is the mid-infrared dominant and only feature. Its appearance correlates with the localized screening of the highest vibrational mode reststrahlen band. We propose that the longitudinal optical mode macroscopic field screening is consequence of dynamically sharing δ disproportioned $e_g$ electrons hovering over the JT distorted octahedral dimer $[Mn\ (Q_{JT})^{3+\delta}\ (Mn(Q_{JT})^{3-\delta}))O_{6/2}]_2$. A thermal driven insulator-metal transition is detected with onset ~1600 K. We also address the occurrence of an inhomogeneity induced THz band result of heating the samples in dry air, triggering $Mn^{3+}$-$Mn^{4+}$ double exchange, under the presence of $Mn^{4+}$ smaller ions stabilizing the orthorhombic lattice.

# Introduction

The nature of distorted perovskite oxides has been a subject of intense scrutiny during last decades owing to competing mechanisms present in the interplay of lattice- charge- spin degrees of freedom providing grounds to elucidate fundamental developments on technologically relevant novel materials. In these systems, electric dipoles, and magnetic moments may couple fulfilling the primary requirement for magnetoelectrics in which ferroelectricity and magnetism coexist in the so-called multiferroic phases.[1]

The structural and dynamical properties of multiferroic perovskites at high temperatures have been only addressed in recent years. Detailed manganite structural temperature induced distortions and related phenomena above ambient temperature were measured in hexagonal $TmMnO_3$ (**Refs. 2, 3**) and in orthorhombic $NdMnO_3$ (**Refs. 4, 5, 6**). The $Mn^{3+}$ allowed $Q_3$ and $Q_2$ JT distortions competing with octahedral rotations is characteristic of the so-called O´ topology sharing strong cooperative Jahn-Teller distortion of the $MO_6$ octahedra inducing orbital ordering. It entails rotation of the Jahn-Teller (JT) ion with a single electron with orbital degeneracy $Mn^{3+}(t_{2g}^3 eg^1)$, responsible of the O' phase, superposed to the JT cooperative distortion short and long Mn-O distance alternation in the **ab** plane.[4, 5] It was found in $NdMnO_3$, above ~973 K, a phase coexisting region with two substructures, the so called O-orthorhombic (octahedral cooperative buckling) and the O'-orthorhombic (octahedral cooperative buckling and JT distortions) in which an orbital disordered two-phase coexist.[6, 7] At still higher temperatures, a JT ordering distortion of $MnO_6$ octahedra goes undetected suggesting a metrically net cubic lattice underlying a possible dynamical JT effect.[8] The interval for J-T ordering and short-range fluctuations has also been reported for PrMnO$_3$ **Ref (9)** and for SmMnO$_3$ **Ref. (10).** The transition temperature increases with the rare earth atomic number.[9]

It is also known that the Rare Earth (R) ionic radius influences directly lattice, orbital, and magnetic properties. Smaller radius makes more competitive the non-perovskite hexagonal phase turning metastable the orthorhombic phase by distortion of the A perovskite cage. ErMnO$_3$ is one of these compounds where the mismatch R-O and Mn-O naturally drives, by octahedral tilting increasing M-O-M angles and shortening Mn-O bond lengths, the more stable non-perovskite hexagonal structure (*P6$_3$cm*). It was found, nevertheless, that the orthorhombic phase can also be prepared by using low-temperature soft chemistry procedures.[11]

Thus, aiming to reduce the existing void on optical properties in RMnO$_3$ compounds at high temperatures we present here far- and mid-infrared emission and reflectivity measurements of distorted perovskite O-ErMnO$_3$ spanning from 40 cm$^{-1}$ to 13000 cm$^{-1}$ and increasing the temperature up to sample dissociation.

At low temperatures, orthorhombic O-ErMnO$_3$ becomes antiferromagnetic correlated below 42 K, locking-in at 28 K but remaining low temperature incommensurate. We did not find phonon profile changes in the temperature range from 12 K to 300 K suggesting phonon band splits leading to an incommensurate K$_2$SeO$_4$ -like lattice and a lock-in transition **Ref (12)** and improper ferroelectricity as proposed by Goto et al for TbMnO$_3$ and DyMnO$_3$ below T$_N$.[13] On cooling at ~40 K, and in full agreement with the reported by Ye et al,[14] we did observe at magnon energies a zone center spin resonance that accordingly is assigned to the emergence of short range spin paramagnon correlations.

Another set of weak features in the lower frequency region have profiles reminiscent to excitations discussed by Kadlec et al.[15, 16] In our case, below 140 cm$^{-1}$, these may be even followed in the far infrared region merging into the background up close to the high temperature order-disorder orbital transition. They might provide grounds for the origin of a local polarization of magnetic modulations[17] contributing to small polaron hopping in the O'-O orthorhombic interphase.

After verifying that the number of normal modes at 12K is close to the predicted for the orthorhombic space group, we found that our reflectivity spectra in the Jahn-Teller orbital ordered phase perfectly match those from emissivity. Both techniques show phonons in the far infrared, where sublattices O-O' coexist orbital disorderly, increasing softening and broadening, and merging up to a temperature in which the phonon spectrum undergoes a rather abrupt change. This signals a first order order-disorder phase transition passage into cubic symmetry at $T_{cubic}$ ~1329 K ± 20 K above which the number of infrared active modes coincides with the three predicted for the space group Pm-3m (Z=1).[18,19,20]

Since the concerted knowledge of temperature evolution of lattice distortions and consequent orbital rearrangement is key in identifying O-ErMnO$_3$ intrinsic properties, we then monitored the consequences of electron-phonon interactions in the mid-infrared. Strong electron-phonon interactions in 3d-2p hybridized orbitals in oxides is a main subject exploiting the strong anisotropic deformability of oxygen ions. It engulfs the so-called polaronic effects in transport, magnetic, and structural properties.[21,22] The fingerprint of such interactions are mid-infrared spectra associated to photoexcited bands the origin of which are quasiparticle absorptions due to self-trapped charges in lattice deformations. These quasiparticles are named polarons and bipolarons.[23] We found that the small polaron view holds at temperatures for which dilution of the long range Jahn-Teller ordered phase into the orbital disordered takes place. This change is accompanied, starting at T* ~700 K in ErMnO$_3$, and as already pointed above, by progressive softening and reduction in the number of vibrational bands. Those mid-infrared spectra are analyzed using the temperature dependence of the optical conductivity using Riek's small polaron proposition.[24]

At $T_{cubic}$ ~1329 K a remarkable well defined mid-infrared band appears centered ~5800 cm$^{-1}$ that, after Reik's small polaron energy and broadening renormalization, is identified due to small bipolarons.[25] Two hundred and fifty degrees further up in temperature the bipolarons peak position softens transferring weigh to an emerging Drude correlated carrier continuum considered as a thermal activated onset of an Anderson-like transition. We then propose that the bipolaron be a dimer composed by two JT distorted dynamically coupled octahedra.

At ~1670 K, we found that the emission spectra helps to corroborate the overall polaron identification and behavior. A bipolaron is destroyed by the photon creating one electron polaron, either at a favorable new lattice site, or contributing to the correlated infrared continuum in a thermally rarefied environment. The lattice deforms accordingly. This mechanism leaves the original companion as a single quasiparticle at a small polaron site. We verified that when the bipolaron peak frequency softens the remaining background could be reproduced using the same arguments for small polarons in the O'-O phases. That is, we corroborate that the photon dissociates the bipolaron into small polaron building components in a coexisting process by which thermal driven freer electrons contribute to the Drude tail triggering a high temperature driven insulator-metal phase transition.

## Experimental details

O-ErMnO$_3$ is a highly distorted perovskite insulator borderline between compounds belonging to the orthorhombic space group ($D^{16}_{2h}$-Z=4) and those manganites with smaller rare earth size, adopting the hexagonal structure ($C^3_{6v}$-

Z=6). It is an example of ion size leading to an increase of the lattice distortion thus increasing the spontaneous orthorhombic strain and closing of the Mn-O-Mn angle. In this phase, its distinctive lattice feature is the parameter ratio **c**/√2 < **a** < **b** different from what it is found in normal perovskites where **c**/√2 lies between **a** and **b**.[11]

O-ErMnO$_3$ was prepared from citrate precursors choosing stoichiometric amounts of Er$_2$O$_3$ and MnCO$_3$ dissolved in citric acid to which it was added few droplets of HNO$_3$. The resulting solution was slowly evaporated leading to a resin first dried at 120º C and then decomposed at 600ºC. A subsequent treatment at 700º C in an inert atmosphere eliminates organic material, nitrates, and Mn$^{4+}$ in the resulting product. For preventing the presence of the hexagonal phase in the O-ErMnO$_3$ product special extra emphasis has been put in the amorphous product of decomposition of the citrate at 700º; the atmosphere of O$_2$ (1 atm) used for decomposition, and the heating rate designed to optimize the orthorhombic phase.[11] Fig.1 shows its room temperature high resolution neutron powder diffraction. The refined unit cell parameters at room temperature are a=5.2262(2), b=5.7932(3), c=7.3486(3) Å in good agreement with published data.[11]

It is worth noting, being O-ErMnO$_3$ increasingly metastable, that in the synthesis of O-ErMnO$_3$ there is always a small amount of Mn$^{4+}$ ions considered intrinsic impurities. It helps reducing the octahedral distortion and brings extra stability into the lattice because Mn$^{4+}$ smaller size. It may also promote the implicit M$^{3+}$-Mn$^{4+}$ double exchange. In addition, one has to consider that at the highest temperatures heating samples in dry air may induce additional triggering of double exchange conductivity by Mn$^{3+}$ and Mn$^{4+}$ ions. Known from annealing oxides in air, the freer electrons are consequence of the oxidation process Mn$^{3+}$→Mn$^{4+}$ + 1e- which is equivalent to divalent A$^{2+}$ substitution in a higher-mobility-high-temperature small polaron environment.[6, 11]

Infrared spectra were taken on heating using two experimental facilities, one corresponds to reflection and a second one for emission. From 12 K to room temperature and from 300 K up to about ~850 K we have measured reflectivity with two Fourier transform infrared spectrometers, a Bruker 66 v/s, and a Bruker 113V, respectively, using conventional near normal incidence geometries. A liquid He cooled bolometer and a deuterated triglycine sulfate pyroelectric bolometer (DTGS) were employed to completely cover the spectral range of interest. A gold mirror was used as 100% reflectivity reference. Low temperature runs were made with the sample mounted in the cold finger of a Displex closed cycle He refrigerator.

For high temperature reflectivity we used a heating plate adapted to the near normal reflectivity attachment in the Bruker 113v vacuum chamber. In this temperature range, the spurious infrared signal introduced by the hot sample thermal radiation was corrected to obtain the absolute reflectivity values.

The reported absorbance measurements have been done at the THz beamline at the electron storage ring BESSY II in the Helmholtz-Zentrum Berlin (HZB) using the low-alpha mode. In the synchrotron low-alpha mode electrons are compressed within shorter bunches of only ~2 ps duration allowing far-infrared wave trains up to mW average power to overlap coherently in the THz range below 50 cm$^{-1}$. We use a superconducting magnet (Oxford Spectromag 4000, -10 T to +10 T) interfaced with the interferometer for the measurements under magnetic fields. Some were also taken in the zero field cooled mode using an Oxford Optistat cryostat in the sample compartment of the IFS125 HR.

The reported absorbance spectra were normalized using as reference the same pellet transmission at temperatures at which there was only a flat response. The temperature was measured with a calibrated Cernox Sensor from LakeShore Cryotronics mounted to the cooper block that holds the sample in the Variable Temperature Insert (VTI) of the Spectromag 4000 Magnet.[26]

Normal emissivity was measured with two Fourier transform infrared spectrometers, Bruker Vertex 80v, and Bruker Vertex 70, coupled to a rotating table placed inside a dry air box allowing to simultaneously measure the spectral emittance in two dissimilar spectral ranges from 40 cm$^{-1}$ to 13000 cm$^{-1}$.

The sample, which is heated with a 500 W pulse Coherent $CO_2$ laser, was positioned on the rotating table at the focal point of both spectrometers in a position equivalent to that of the internal radiation sources inside the spectrometers. In this measuring configuration, the sample, placed outside the spectrometer, is the infrared radiation source, and conversely, the sample chamber inside the spectrometers is empty. Note that at the highest laser heating powers the bottom face of the pellet that is exposed to radiation might be damaged generating extrinsic defects. For this reason, to assure reliability and reproducibility, the shown measurements are result of 10 independent runs each run using a fresh sample between 400 K to ~2400 K.

Even that the experimental set up has been earlier discussed, for the sake of clarity we will now repeat below the basic concepts and tools used in the data analysis.[27, 28,]

Normal spectral emissivity of a sample, $\boldsymbol{E}(\omega,T)$, is given by the ratio of its luminescence ($L_S$) relative to the black body's luminescence ($L_{BB}$) at the same temperature $T$ and geometrical conditions, thus,

$$\boldsymbol{E}(\omega,T) = \frac{L_S(\omega,T)}{L_{BB}(\omega,T)} \qquad (1)$$

In practice, the evaluation of this quantity needs the use of a more complex expression because the measured fluxes are polluted by parasitic radiation. This is because part of the spectrometer and detectors are at 300 K. To eliminate this environmental contribution the sample emissivity is retrieved from three measured interferograms,[28] i.e.

$$E(\omega,T) = \frac{FT(I_S - I_{RT})}{FT(I_{BB} - I_{RT})} \times \frac{\mathscr{P}(T_{BB}) - \mathscr{P}(T_{RT})}{\mathscr{P}(T_S) - \mathscr{P}(T_{RT})} E_{BB} \qquad (2)$$

where *FT* stands for Fourier Transform., and I for measured interferograms i.e., sample, $I_S$; black body, $I_{BB}$; and, environment, $I_{RT}$. $\mathscr{P}$ is the Planck's function taken at different temperatures T; i.e., sample, $T_S$; blackbody, $T_{BB}$; and surroundings, $T_{RT}$. $E_{BB}$ is a correction that corresponds to the normal spectral emissivity of the black body reference (a LaCrO$_3$ Pyrox PY 8 commercial oven) and takes into account its non-ideality.[28]

One of the advantages of emissivity is that allows contact free measurement of the temperature of a measured insulator using the Christiansen frequency; i.e., the frequency where the refraction index is equal to one, and the extinction coefficient is negligible, just after the highest longitudinal optical phonon frequency. The temperature is calculated with emissivity $E(\omega,T)$, eq. (2), set equal to one at the Christiansen frequency. Thus, in a regular run one interferometer always measures the infrared DTGS region, where the Christiansen point lies, while the other covers any other spectral region of interest, both, focused at the same sample spot. This is the reason of having two interferometers measuring simultaneously. In the measurements discussed in this manuscript, in addition to the infrared region covered by a DTGS detector, the spectral regions of interest were the far infrared from 40 cm$^{-1}$ to 900 cm$^{-1}$ using a He cooled bolometer, and another from 7000 cm$^{-1}$ to 13000 cm$^{-1}$ with a high gain InGaAs photodiode.

After acquiring the optical data we placed our spectra in a more familiar near normal reflectivity framework using the second Kirchhoff law, that is,

$$R = 1 - E \qquad (3)$$

where $R$ is the sample reflectivity. It assumes that in the spectral range of interest any possible transmission is negligible being emissivity the complement of reflectivity.

This allows computing oscillator frequencies using a standard multioscillator dielectric simulation[29,30] with the dielectric function, $\varepsilon(\omega)$, given by

$$\varepsilon(\omega) = \varepsilon_1(\omega) - i\varepsilon_2(\omega) = \varepsilon_\infty \prod_j \frac{(\omega_{jLO}^2 - \omega^2 + i\gamma_{jLO}\omega)}{(\omega_{jTO}^2 - \omega^2 + i\gamma_{jTO}\omega)} \tag{4}$$

$\varepsilon_\infty$ is the high frequency dielectric constant taking into account electronic contributions; $\omega_{jTO}$ and $\omega_{jLO}$, are the transverse and longitudinal optical mode frequencies and $\gamma_{jTO}$ and $\gamma_{jLO}$ their respective damping. We also added when needed a double damping extended Drude term (correlated plasma contribution) to the dielectric simulation,

$$-\frac{(\omega_{pl}^2 + i \cdot (\gamma_{pl} - \gamma_0) \cdot \omega)}{(\omega - i\gamma_0) \cdot \omega} \tag{5}$$

where $\omega_{pl} = \sqrt{\frac{n e^2}{\varepsilon_0 \varepsilon_\infty m^*}}$ is the plasma frequency (*n*, the number of carriers, *e*, the carrier charge; *m\**, the effective mass). $\gamma_{pl}$ its damping, and $\gamma_0$ (the inverse of relaxation time at zero frequency) understood as a phenomenological damping introduced to reflect lattice drag effects. When these two dampings are set equal, one retrieves the classical Drude formula.[31]

When the Christiansen point becomes a less reliable temperature reference, as in poor conducting samples where a Dudre term may conceal or distort the actual frequency minimum, we imposed as extra condition that sample emission cannot be higher than one.

The real $(\varepsilon_1(\omega))$ and imaginary $(\varepsilon_2(\omega))$ part of the dielectric function (complex permittivity, $\varepsilon^*(\omega)$) is then estimated from fitting the data using the reflectivity R given by[32]

$$R(\omega) = \left| \frac{\sqrt{\varepsilon^*(\omega)} - 1}{\sqrt{\varepsilon^*(\omega)} + 1} \right|^2 \qquad (6)$$

We also calculated the real part of the temperature dependent optical conductivity, $\sigma_1(\omega)$ given by[32]

$$\sigma_1(\omega) = \frac{\omega \cdot \varepsilon_2(\omega)}{4\pi} \qquad (7)$$

This constitutes our measured optical conductivity.

## RESULTS AND DISCUSSION

### i) *Phonons, magnetic excitations, and small polarons in the orthorhombic phase*

In order to study O-ErMnO$_3$ at the highest temperatures we first supported our work with measurements of near normal reflectivity in the orthorhombic phase from 12 K up to ~ 800 K. Fig. 2(a) shows the 12 K phonon reflectivity spectra and multioscillator fit from which it is possible to identify 21 of the 25 vibrational bands (Table I) predicted by group theory for the orthorhombic space group P*bnm*-$D^{16}_{2h}$-(Z=4)

$$\Gamma_{IR}(O') = 7B_{1u} + 9B_{2u} + 9B_{3u} \qquad (8)$$

At low temperatures we also find in the THz region a distinctive convoluted band centered at ~45 cm$^{-1}$ (Fig. 3a, inset) assigned to the Kramers doublet of odd $f$ electron Er$^{3+}$(4f$^{11}$). Under low applied magnetic fields each component of the doublet undergoes a weak splitting suggesting a linear field dependence.(Fig. 3 (b), inset)

At still lower frequencies we find a band which origin is not totally understood. Its temperature dependent reflectivity spectra may simply be reproduced by fitting to a Gaussian line shape (Fig. 2(a), inset (b)); Table I)).

O-ErMnO$_3$ is known to have an amount of M$^{4+}$ ions, estimated in our case ~5%,[11] that are considered intrinsic impurities, and that in addition to oxygen vacancies, turn this compound into a partially compositional disordered material. The random lattice disorder is associated to structural inhomogeneity, unreleased strains, and potential freer charges localized in grain boundaries or domain walls that cause, in defective crystals, at lower than phonon frequencies the so-called boson peak. Vacancies and interstitials act as local scattering centers that may relate to some structural disorder at atomic level. The local structure inhomogeneity has been also invoked to model the boson peak as function of randomly-cut bonds and local breaking of the center inversion symmetry with any atom taken as the measured local center symmetry.[33]

We identify the band centered $\omega_{0G}$~20 cm$^{-1}$ at 12 K (Table 1) as arising from structural inhomogeneities yielding a collective temperature dependent feature that harden and weakens on cooling. Its full width at half-maximum increases with temperature.

The small amount of impurities and vacancies may also trigger a scenario by which ferroelectricity would be of the relaxor type, as it was found in isomorph low doped TbMnO$_3$ (Ref. 34) related to local order competition.[35] It also means on having the possibility of finding local polarization above T$_c$ at paraelectric-ferroelectric transition even in an environment lacking of macroscopic spontaneous polarization and structural symmetry breaking as in conventional ferroelectricity. In this scenario, high temperature dipolar fluctuations of paraelectric domains would slow down on cooling creating nanoregions made of randomly oriented electric dipoles that eventually freezes.[36] It may then be associated with temperature dependent behavior of the band tail found in O-ErMnO$_3$ (Fig. 2(a), inset (b)). From a very broad profile at room temperature reduces to a narrower and hardened weaker bell shape at 12 K. That is, although O-ErMnO$_3$ is nominally centrosymmetric Pbnm, with ferroelectricity expected from an spiral magnetic structure breaking time reversal and inversion symmetry[37] the small amount of Mn$^{4+}$ ions and oxygen vacancies of real life O-ErMnO$_3$, diminishes the coherence of the spin-spiral structure as in TbMnO$_3$ (Ref. 34). The unsettling presence of Mn$^{4+}$ suggest why only short range magnetic order is found in O-ErMnO$_3$.[14] Mn$^{4+}$ ions alter the Mn-O-Mn angle key in determining the exchange coupling,[38] thus disrupting the spin spiral, and consequently, suppressing magnetically driven ferroelectricity, turning O-ErMnO$_3$ into a just incipient multiferroic.

On the other hand, it is also worth noting that transmission measurements show that those electric dipole homogeneities (Fig. 2(a), inset (b)) coexist with a weak magnetic counterpart measured at the higher intensity far infrared beam of the alpha mode at the THz beamline in BESSY II. Figure 3(a) shows that in addition to Er$^{3+}$ crystal field transition there appears another well defined but several orders weaker band emerging at ~40 K. This is in full agreement with the temperature of the short range antiferromagnetic correlated paragmagnon reported by **Ref (14).** Its magnetic nature is revealed under applied fields and what it is more interesting, it shows that the original profile distorts easily into a much broader featureless band under relatively small fields. This finding is likely due to the reported absence of long range magnetic correlations and in contrast with our earlier measurements in orthorhombic ErCrO$_3$ where the expected set, the ferromagnetic and antiferromagnetic spin wave resonances, are well defined.[39]

Above ~40 K, magnetically uncorrelated ions may still play a magnetic subtle role being spin and charge two fundamental degrees of freedom in multiferroics. Then, not limited to the low temperature range, these random oriented magnetic moments in the not randomly coupled insulator may help to consolidate energetically favorable quasiparticles in the collective interplay of the solid. Electrons, along with a distortion of the surrounding lattice site, and as a result of the strong ion interaction within the lattice and possible unaccounted magnetic correlations, may yield new bound entities as those identified in paragraphs of this paper in the crystalline order of high temperature cubic ErMnO$_3$.

As said, the room-temperature orthorhombic lattice is built by the so called O-orthorhombic topology with cooperative buckling of the corned shared octahedral. Increasing temperature gives place to the O'-orthorhombic phase only found when the cooperative J-T distortion is superposed to the rotation of the MnO$_6$ octahedra in the O-orthorhombic phase.[7] The coexisting O'-O orthorhombic phases translates in a continuous reduction in the number of bands by softening and merging, shown in Fig. 2(b) for high temperature reflectivity and emissivity, that it is also in consonance with an underlying thermal driven dynamical distortions of the Jahn-Teller octahedras into a metric cubic environment.

The complete overall temperature evolution of ensuing phonons in the far infrared is shown in Fig 4(a)). At about 1329 K ± 20 K there is a rather abrupt change in profiles assigned to the structural transition into the high temperature cubic phase. This distinctive behavior at T$_{cubic}$, is also linked to emerging bands in the 1-E mid-infrared spectra, Fig. 4(b), consequence of the temperature dependent orbital rearrangement.

In the following paragraphs, we detail far infrared spectra below, at, and above the structural phase transition, to then continue with the temperature dependent analyses of their corresponding optical conductivity $\sigma_1(\omega)$.

At 882 K, and 1255 K, the number of oscillators needed to satisfy fits of the far infrared phonon activity of O-ErMnO$_3$ is only 6 (Table II) pointing, as in NdMnO$_3$ **(Ref. 6),** to the increase in the net lattice symmetry through thermal fluctuations mirrored in broadening asymmetries mainly associated with octahedral internal modes. At these temperatures the boson peak, Fig. 2((a), inset(b)) becomes a much broad overdamped band mostly outside of our lower frequency range of detection. For this reason, knowing that in some instances has also been associated to a lowest phonon-like van-Hove singularity,[40] we simulated the sloped background below ~60 cm$^{-1}$ with an arbitrary extra oscillator in addition to those modeling phonons (tables (II, III)).

Lattice modes, arrows in Fig. 5(a, b), merge without complete mode softening when getting closer to the structural transition as in an order-disorder process.

In order to achieve a better understanding of the high temperature processes we should also now note that in addition of line broadening and number reduction there appears a band centered at ~1800 cm$^{-1}$ associated with cooperative stripe-ordered small JT polarons at 882 K.[41] According to the known sequence in NdMnO$_3$,[4,5] by increasing the temperature the orbitals gradually mix disorderly up to the equivalent of the so-called T$_{JT}$ temperature above which the cubic symmetry prevails. This results in the mid-infrared smoother plateau at 1255 K (Fig. 4b). In the emissivity nomenclature, it is said that the plateau is consequence of the sample higher opacity.

The mid-infrared feature at ~1800 cm$^{-1}$ is known in materials associated with bound quasi-particles, small polarons, in which the oxygen polarizability, and thus, distorted orbitals in metastable lattices, play a basic role. A small polaron is made of an electron laying in the ion potential created by the disturbance that the charge carrier creates.[42] Its optical detection is due to a self-trapped carrier excited from its localized stated to a localized state at a site adjacent to the small polaron site.[43] Its range is usually less than the unit cell size having as main transport property a characteristic thermal activated hopping accompanied by the lattice deformation. The implied scenario is replicated in systems than range from oxides to polymers and has been the subject of many reviews.[25, 44, 45]

We extract the parameters characterizing these quasiparticles from our fits to the experimental optical conductivity real part $\sigma_1(\omega)$, eq.(7), using the theoretical formulation by Reik and Heese.[24,46] Small polarons are addressed microscopically as due to nondiagonal phonon transitions in materials having a sublattice made of oxygen ions as common denominator. The optical conductivity is calculated for carriers in one small band and interband transitions, also detected in our spectra, are excluded. Starting with a Holstein's Hamiltonian,[47] the frequency dependent conductivity is calculated using Kubo's formula.[48,49]

Then, the real part of the model optical conductivity for a finite temperature T, $\sigma_1(\omega,\beta)$ is given by

$$\sigma_1(\omega,\beta) = \sigma_{DC} \frac{\sinh\left(\frac{1}{2}\hbar\omega\beta\right)\exp\left[-\omega^2\psi^2 r(\omega)\right]}{\frac{1}{2}\hbar\omega\beta\left[1+(\omega\psi\Delta)^2\right]^{1/4}}, \qquad (9)$$

$$r(\omega) = \left(\frac{2}{\omega\Psi\Delta}\right)\ln\left\{\omega\Psi\Delta + \left[1+(\omega\Psi\Delta)^2\right]^{1/2}\right\} - \left[\frac{2}{(\omega\Psi\Delta)^2}\right]\left\{\left[1+(\omega\Psi\Delta)^2\right]^{1/2} - 1\right\}, \quad (10)$$

with $\Delta = 2\varpi_j \Psi$ (11)

and $\Psi^2 = \dfrac{\left[\sinh\left(\dfrac{1}{2}\hbar\varpi_j \cdot \beta\right)\right]}{2\varpi_j^2 \eta}$. (12)

The most useful point of this approach is that it is built on data acquired separately in independent measurements allowing a self-check approach for the fits to the experimental optical conductivity. The model conductivity, $\sigma_1(\omega,\beta)$, $\beta=1/kT$, is only three parameter dependent; $\sigma_{DC}=\sigma(0,\beta)$, the electrical DC conductivity; the frequency $\varpi_j$ that corresponds to the average between the transverse and the longitudinal optical mode of the $j^{th}$ restrahlen band; and $\eta$, is a parameter characterizing the strength of the electron-phonon interaction, i.e., the average number of virtual phonons contributing to the polarization around a localized polaron. It is also proportional to the small polaron binding energy, $E_b = \dfrac{\eta \varpi_j}{2}$. From them, $\eta$, is the only truly free parameter in the computation for each phonon $\varpi_j$. This is because phonon frequencies, $\varpi_j$, are fixed by the reflectivity (or 1-emissivity), known from the independent fits, (table II), and the $\sigma_{DC}=\sigma(0,\beta)$-zero frequency-conductivity is known from either optical or transport measurements.[50] $\eta \sim 3$ implies a low to mild electron-phonon interaction while a value around 14 or higher would correspond to the very strong end.[51]

Our measured and calculated optical conductivity at 882 K and that for the intermediate JT orbital disordered phase at 1255 K are shown in Fig. 5(a, b; lower panels). To reproduce the mid-infrared spectral region entirely, we allowed the possibility of more than one vibrational active contribution using an empirical approach shown successful when applied to other distorted perovskites and glassy systems.[6, 52, 53] This results in the convoluted addition of two bell shaped uncorrelated individual contributions, each of them, calculated at a phonon frequency $\varpi_j$ and a temperature T , in the region from 1000 cm$^{-1}$ to 7000 cm$^{-1}$. Our fit at 882 K confirms the association of the spectral distinctive bump at ~1800 cm$^{-1}$, with the cooperative short-range octahedral Jahn-Teller orbital distortion product of symmetric stretch octahedral vibrational mode, which is expected to play a main role in strong electron-phonon interactions. Our $\eta$´s (table IV) are for strong electron phonon interactions. This picture varies little for 1255 K, and it is expected that remains valid up to the temperature of the structural change to the next discussed cubic phase.

*ii)*   *Phonons and small bipolarons in the cubic phase*

Fig 6 (a,b) and Table III show infrared active phonons in the cubic phase close to the phase transition. Their number agrees with twelve zone center optical phonons expected for the triple degenerate irreducible representation Γ= 3F$_{1u}$ + 1 F$_{2u}$ in the high temperature cubic space group P*m-3m* (Z=1) (pseudocubic P*mcm*, Z=1). Increasing symmetry, there is lattice mode merging (arrows in Figs. 5, 6) and, in contrast to the orthorhombic, a locally dampened restrahlen of the breathing mode at ~600 cm$^{-1}$. (red circle; see band contrasting behavior in Fig. 5 against Fig. 6) which anomaly recalls local fluctuations associated with the short-range polaron of a Jahn Teller distortion found by inelastic neutron scattering approaching the metal-insulator transition temperature in La$_{0.7}$Ca$_{0.3}$MnO$_3$.[54]

At $T_{cubic}$ interacting electrons and phonons pass entangled into the cubic phase through the order-disorder phase transition. In the cubic phase (the would be paraelectric in a conventional soft mode paraelectric-ferroelectric transition)[55] there is no trace of a zone center soft mode but it is possible to infer, linked to the existence of bipolarons over at least 200 degrees above $T_{cubic}$, local off center distortions.[56] This distortions contribute to the overall increment in the vibrational band half-width at half-maximum in an environment with thermal fluctuations so strong that increasing the temperature will increase thermally driven carrier mobility preventing the observation sharper substructure at phonon frequencies.

The zone center $F_{1u}$ phonons are identified as the lattice modes by which Er ions move against oxygen octahedra, Mn ions against the plane oxygens-bending mode, and the octahedral stretching-breathing mode. A $F_{2u}$ fourth involving a scissor motion of the in-plane oxygens is silent.[18, 19, 20]

We also allow a low frequency Drude contribution eq. (5) in the analyses of all spectra taken above $T_{cubic}$ to assure continuity in the high temperature fits. While a very weak Drude term helps in reproducing the low frequency 1-E slope, in the semilog plots there is a more meaningful contribution increasing temperatures. Below ~1600 K, (Fig. 6 (a, b)), freer electrons, have still an extremely high effective mass nullifying any significant role.

In addition, and no less distinctive, is the strong broad band emerging in the mid-infrared response (Fig. 4 and Figs. 6 (a),(b) inset) characteristic of the cubic phase and contrasting against the near flat orthorhombic response below 1329 K (Fig. 4 and Figs 5 (a),(b) inset). We identify the band centered at ~5800 cm$^{-1}$ as due to a small bipolaron, a localized quasiparticle made by a bound state of two electron small polarons paired in real space.[23] Band profiles and energies closely follow similar features found by Puchkov et al[57] and others[25] in mix-oxides.

We use a simplified version for the optical conductivity also proposed by Reik, that retaining the basic concept of the small polaron original proposition, supplies the absence of the microscopic approach. That is,

$$\sigma(\omega, T) = \sigma_{DC} \frac{sinh(4E_b'\hbar\omega/\Delta'^2)}{4E_b'\hbar\omega/\Delta'^2} exp(-(\hbar\omega)^2/\Delta^2), \quad (13)$$

where $\sigma_{DC}=\sigma(0,\beta)$ is, as before, is the electrical or optical DC conductivity.

Here it is necessary to consider that the confinement energy of small bipolarons is double relative to that in small polarons because the presence of the second electron, i.e., the depth of the well that self-traps both carriers is twice as deep. The electron-phonon coupling energies are quadrupled, and at the same time, both charged carriers repel through their U Coulomb interaction, i.e., the bipolaron is stable if $2E_b>U$. Accordingly, peaking of a band associated with bipolarons will be found at higher frequencies than for small polarons.[42] This means that eq(13) describes localized small bipolarons when the $2E_b$, the peak energy of the small polaron absorption, is replaced by $2E_b'=4E_b-U$, the experimentally measured absorption maximum as in eq (13), and $\Delta$ by $\Delta'= \sqrt{2E_b'E_{vib}}$, where $E_{vib}$ corresponds to a relevant phonon.[25] In our case, the highest frequency longitudinal optical mode at ~640 cm$^{-1}$. The strong electron-phonon interaction by the octahedral internal mode has as a net result electrons overwhelming their Coulomb repulsion energy U to form bipolarons.[58] The bipolaron effective mass is expected to be also much larger than the hundred fold increments of the electron mass expected for small polarons.

From ~ 1350 K to ~ 1600 K, (Figs. 6, 7, 8) our spectra suggest no structural changes. The only differences among the spectra is the far infrared appearance of a Drude component which does not considerably grows in intensity, together with the anomalous steady increment in the breathing mode reststhalen smearing at ~600 cm$^{-1}$. The lower panels show the very good agreement achieved comparing the experimental optical conductivity from the fit against those calculations using peak position and zone center DC conductivity as free parameters in eq(13).

Approaching T ~ 1630 K, (Fig. 8(a)) thermal activation and "resonant" radiation become more relevant. Photon induced, one of the pair dimerized trapped electrons constituting the bipolaron hops to a next near neighbor small polaron site. This hopping motion, to which thermal fluctuation should also be added, increases conductivity by the dissociated $e_g$ electrons prompting an incipient insulator into a poor conducting solid. At ~1630 K, there is relative weaker overdamped zero frequency centered Drude contribution and its corresponding tailing (Fig, 8 a, table III). It offers a quantitative picture on the emerging freer carrier activity that spreading in a fairly broad range determines a thermal driven insulator-metal phase transition. And, as pointed already above, it also reduces the usefulness of the Christiansen point as quantitative thermometer.

Increasing the temperature further, Fig 8 (b), the Drude term grows distinctly. If the carrier's effective mass is assumed constant the ratio of the plasma frequency at the two temperatures, eq. (5), yields an increment of the ∼24 % in the number of freer carriers. At ~1670 K the bipolaron band peak position softens, yielding a mid-infrared spectrum in which the bipolaron profile at 3000 cm$^{-1}$ is followed by an undistorted tail up to 10000 cm$^{-1}$ (Fig. 8 (b), inset). The tail is close to that found in the mid-infrared response of small polaron in orthorhombic phase what suggests an interpretation involving localization for the dimer quasiparticles and an intermediate picture reflecting single polaron optical absorption in a many body polaron scenario. The lower panel of Fig. 8(b) shows that the shape of the mid-infrared conductivity at ∼1670 K may be indeed be reproduced by plain superposition of the two contributions, (eqs. 9 and 13)) i.e, localized bipolarons and small polarons, product of the bipolaron dissociation, coexisting with thermal activated itinerant carriers.

### iii) A proposition toward a bipolaron model

The case for bipolarons in oxides goes back to a proposal by Sugai et al for low temperature high Tc $Ba_xK_{1-x}BiO_3$ and $BaPb_{1-x}Bi_xO_3$ mixed perovskites.[59,60] These also share an orthorhombic to cubic phase transition in which bipolarons dimers, as in $[Bi(Pb)O_{6/2}]_2$, are identified. The idea is based on an earlier work by Rice on overlapping electrons between adjacent molecular sites and Coulomb interactions between radical electrons.[61]

We propose that, in cubic $ErMnO_3$, bipolarons are a consequence of a JT dynamic arrangement in which crystal field split is dimer compromised.

Summing up as background the orbital short-long JT M-O ordering in the orthorhombic *__ab__* plane, and surmounting the orbital disordered intermediate phase, diamagnetic dimers made of two octahedras are proposed making a molecule as $[Mn(Q_{3JT})^3(Mn(Q_{3JT})^3))O_{6/2}]_2$. $Q_{3JT}$ is identified with the distortion oxygen displacements corresponding to the filled $3r^2-z^2$ and $x^2-y^2$ orbitals yielding the cooperative JT deformation in the orthorhombic *__ab__* plane (at this stage we do not exclude a possible role for the $Q_2$ distortion stabilizing both orbitals).[4] Here, electrons wave function is dynamically protracted over the dimerized molecule as in $[Mn(Q_{3JT})^{3+\delta}(Mn(Q_{3JT})^{3-\delta}))O_{6/2}]_2$ -being $\delta$ understood as a dynamical net charge disconmensuration needed to explain the anomalous local macroscopic field associated with the longitudinal stretching mode (blurred restrahlen band in Figures 6, 7, 8 at ~600 cm$^{-1}$). At 1670 K its trace disappears completely (Fig. (8 (b)). The situation is reminiscent to "charge – oscillation" mixing of the unpaired electrons of pairs of molecular ion radicals coupled by the molecular vibration.[62]

The well-defined bipolaron profiles in the 1400 K to 1600 K may then be though as a result of bipolaron condensation from a high temperature continuum into an ordered state. And finally, that scenario may also include a fluctuating valence configuration $[Mn^{4+}(Mn(JT)^{3+}))O_{6/2}]_2$, present if impurity $Mn^{4+}$ ions are taken into account, i. e., $Mn^{4+}$ ions originating in extra oxygens as extrinsic factor that being smaller than $Mn^{3+}$ help to stabilize the orthorhombic lattice.[11]

## CONCLUSIONS

We report, using reflection and emission techniques in the far- and mid-infrared, on the temperature structural and phonon evolution of ambient orthorhombic O-ErMnO$_3$. We found that between 2 K and 12 K the space group allowed phonons to coexist with a paramagnon spin resonance and Rare Earth crystal field transitions. Increasing the temperature a number of vibrational bands undergo profile broadening and softening approaching the orbital disordered phase where the orthorhombic O´ lower temperature cooperative phase coexists with cubic-orthorhombic O. O-ErMnO$_3$ undergoes a first order order-disorder transition into the perovskite cubic phase at $T_{cubic}$ ~1329 K ± 20 K. This is identified spectroscopically by the number of phonons coincident with predictions by the space group Pm-3m (Z=1).

In addition to the vibrational behavior, we found in the orthorhombic phase small polarons with strong electron-phonon interactions linked to antisymmetric and symmetric stretching modes. This strong electron-phonon interactions make polaronic carriers bound into almost immobile bipolarons above $T_{cubic}$,[63] being their profile in the mid-infrared a distinctive feature.

We propose that our data may be understood by the formation of condensed bipolarons made of coupled JT-distorted octahedra, as in dimer $[Mn(Q_{3JT})^{3+\delta}(Mn(Q_{3JT})^{3-\delta}))O_{6/2}]_2$ -being $\delta$ disproportioned $e_g$ electrons dynamically sharing and screening the macroscopic field associated with the longitudinal optical phonon at ~640 cm$^{-1}$. Increasing further the temperature triggers an insulator metal transition at ~1600 K recalling earlier predictions on the bipolaron role in high Tc oxides.[64]

We also address the occurrence of an inhomogeneity induced band at THz energies as possible consequence of heating the samples in dry air, triggering $Mn^{3+}$-$Mn^{4+}$ double exchange, under the presence of a minor amount of $Mn^{4+}$ smaller ions stabilizing the orthorhombic lattice.

## Acknowledgements

The authors are indebted to D. De Sousa Meneses *(*Conditions Extrêmes et Matériaux: Haute Température et Irradiation - UPR3079 CNRS (C.E.M.H.T.I.)) for sharing his expertise on infrared techniques. NEM is also grateful to the CNRS-C.E.M.H.T.I. laboratory and staff in Orléans, for research and financial support in performing far infrared reflectivity and emissivity. He also thanks BESSYII at the Helmholtz-Zentrum Berlin für Materialien und Energie for beamtime allocation under Project Nº 2013-1-120813 and financial assistance. JAA acknowledges the Paul Scherrer Institute (Swizerland) for the allowed neutron time, and the financial support of the Spanish "Ministerio de Economia y Competitividad" (MINECO) through Project Nº MAT2017-84496-R.

# TABLE I

Dielectric simulation fitting parameters for orthorhombic ErMnO$_3$ at 12 K

(Bottom cells show the parameters used in the lower frequency Gaussian fit at different temperatures shown in Fig. 2 (a), inset (b))

| T (K) | $\varepsilon_\infty$ | $\omega_{TO}$ (cm$^{-1}$) | $\Gamma_{TO}$ (cm$^{-1}$) | $\omega_{LO}$ (cm$^{-1}$) | $\Gamma_{LO}$ (cm$^{-1}$) |
|---|---|---|---|---|---|
| 12 | 2.36 | *33.4* | *38.8* | *38.0* | *18.1* |
|  |  | *68.5* | *17.6* | *73.8* | *19.3* |
|  |  | 119.3 | 129.2 | 139.8 | 104.1 |
|  |  | 165.3 | 10.7 | 167.4 | 14.2 |
|  |  | 178.8 | 24.3 | 181.3 | 5.9 |
|  |  | 184.6 | 4.5 | 189.5 | 18.1 |
|  |  | 196.9 | 32.4 | 200.9 | 5.7 |
|  |  | 203.9 | 6.1 | 206.1 | 10.4 |
|  |  | 215.3 | 33.8 | 224.7 | 16.5 |
|  |  | 228.7 | 15.9 | 247.0 | 100.0 |
|  |  | 265.3 | 28.0 | 266.9 | 17.4 |
|  |  | 290.4 | 24.2 | 297.4 | 16.0 |
|  |  | 317.7 | 9.2 | 319.4 | 6.4 |
|  |  | 339.4 | 9.3 | 342.0 | 7.5 |
|  |  | 374.1 | 25.0 | 377.9 | 96.1 |
|  |  | 397.3 | 217.8 | 14.6 | 126.9 |
|  |  | 406.9 | 15.3 | 408.7 | 18.4 |
|  |  | 423.2 | 22.2 | 425.3 | 17.0 |
|  |  | 460.2 | 72.5 | 465.2 | 29.7 |
|  |  | 504.4 | 47.6 | 521.7 | 27.3 |
|  |  | 535.1 | 32.1 | 559.4 | 35.6 |
|  |  | 583.5 | 15.4 | 668.9 | 99.5 |
|  |  | 687.6 | 74.7 | 695.3 | 49.6 |
|  |  | Peak position $\omega_{0G}$ | Bandwidth | A |  |
|  |  | 20.6 | 6.1 | 113.7 |  |
| 300 |  | ~-67.5 | ~52.5 | ~1303 |  |

# TABLE II

Dielectric simulation fitting parameters for high temperature orthorhombic ErMnO$_3$

(First row italics represent the boson peak parametrization (see text) and the last row is the simulation for the small polaron contribution)

| T (K) | $\varepsilon_\infty$ | $\omega_{TO}$ (cm$^{-1}$) | $\Gamma_{TO}$ (cm$^{-1}$) | $\omega_{LO}$ (cm$^{-1}$) | $\Gamma_{LO}$ (cm$^{-1}$) |
|---|---|---|---|---|---|
| 882 | 1.04 | *61.6* | *37.9* | *66.9* | *53.3* |
| | | 169.8 | 42.9 | 183.7 | 52.5 |
| | | 243.9 | 60.7 | 256.7 | 69.2 |
| | | 380.7 | 71.2 | 404.5 | 143.3 |
| | | 449.9 | 389.9 | 469.9 | 90.7 |
| | | 494.5 | 72.9 | 519.3 | 76.6 |
| | | 555.5 | 95.8 | 723.3 | 82.4 |
| | | *2786.4* | *4509.7* | *3687.6* | *5013.2* |

| T (K) | $\varepsilon_\infty$ | $\omega_{TO}$ (cm$^{-1}$) | $\Gamma_{TO}$ (cm$^{-1}$) | $\omega_{LO}$ (cm$^{-1}$) | $\Gamma_{LO}$ (cm$^{-1}$) |
|---|---|---|---|---|---|
| 1255 | 1.01 | *57.6* | *35.7* | *63.0* | *50.2* |
| | | 166.4 | 45.1 | 179.6 | 50.3 |
| | | 246.2 | 90.7 | 263.6 | 82.8 |
| | | 360.4 | 84.6 | 389.0 | 135.8 |
| | | 444.9 | 389.9 | 471.8 | 104.2 |
| | | 496.5 | 85.2 | 509.9 | 63.6 |
| | | 520.5 | 95.8 | 739.7 | 86.2 |
| | | *1943.7* | *3835.3* | *2415.5* | *4295.0* |

# TABLE III

Dielectric simulation fitting parameters for cubic ErMnO$_3$

(First row italics represent the boson peak parametrization (see text) and the last two rows correspond to the simulation for the bipolaron contribution and interband outline. Note that above ~1500 K we consider that the Drude term overwhelms any other possible contribution at lower frequencies)

| T (K) | $\varepsilon_\infty$ | $\omega_{TO}$ (cm$^{-1}$) | $\Gamma_{TO}$ (cm$^{-1}$) | $\omega_{LO}$ (cm$^{-1}$) | $\Gamma_{LO}$ (cm$^{-1}$) |
|---|---|---|---|---|---|
| 1384 | 1.05 | *104.6* | *889.9* | *155.0* | *292.4* |
|  |  | 160.4 | 86.8 | 211.4 | 178.2 |
|  |  | 354.2 | 154.1 | 375.4 | 177.8 |
|  |  | 558.9 | 185.1 | 639.2 | 205.4 |
|  |  | *5002.2* | *6069.5* | *9054.4* | *3465.5* |
|  |  | *11398.8* | *2963.0* | *11588.9* | *379.9* |
|  |  |  |  | $\omega_{pl}$ (cm$^{-1}$) |  |
|  |  |  | 538.1 | 202.5 | 479.5 |

| T (K) | $\varepsilon_\infty$ | $\omega_{TO}$ (cm$^{-1}$) | $\Gamma_{TO}$ (cm$^{-1}$) | $\omega_{LO}$ (cm$^{-1}$) | $\Gamma_{LO}$ (cm$^{-1}$) |
|---|---|---|---|---|---|
| 1425 | 1.02 | *104.6* | *889.6* | *155.0* | *292.4* |
|  |  | 160.4 | 87.1 | 213.7 | 175.4 |
|  |  | 351.0 | 155.2 | 475.4 | 176.9 |
|  |  | 558.9 | 178.7 | 637.6 | 210.2 |
|  |  | *5008.6* | *6127.7* | *8431.7* | *3117.6* |
|  |  | *10960.8* | *2880.1* | *11589.6* | *271.6* |
|  |  |  |  | $\omega_{pl}$ (cm$^{-1}$) |  |
|  |  |  | 538.1 | 222.5 | 429.5 |

| T (K) | $\varepsilon_\infty$ | $\omega_{TO}$ (cm$^{-1}$) | $\Gamma_{TO}$ (cm$^{-1}$) | $\omega_{LO}$ (cm$^{-1}$) | $\Gamma_{LO}$ (cm$^{-1}$) |
|---|---|---|---|---|---|
| 1523 | 1.03 | 175.8 | 89.1 | 201.9 | 91.9 |
|  |  | 369.1 | 174.1 | 450.8 | 157.3 |
|  |  | 602.4 | 252.3 | 660.5 | 170.0 |
|  |  | *5906.1* | *5236.3* | *9011.1* | *3365.5* |
|  |  | *10769.6* | *1855.3* | *11687.4* | *231.1* |
|  |  |  |  | $\omega_{pl}$ (cm$^{-1}$) |  |
|  |  |  | 613.7 | 443.6 | 578.7 |

| T (K) | $\varepsilon_\infty$ | $\omega_{TO}$ (cm$^{-1}$) | $\Gamma_{TO}$ (cm$^{-1}$) | $\omega_{LO}$ (cm$^{-1}$) | $\Gamma_{LO}$ (cm$^{-1}$) |
|---|---|---|---|---|---|
| 1563 | 1.14 | 177.1 | 98.2 | 205.8 | 87.7 |
| | | 366.1 | 180.1 | 445.8 | 133.5 |
| | | 569.6 | 300.5 | 640.1 | 207.2 |
| | | *6186.1* | *6285.8* | *8767.2* | *4140.5* |
| | | *10809.1* | *767.7* | *11690.1* | *172.4* |
| | | | | $\omega_{pl}$ (cm$^{-1}$) | |
| | | | 576.2 | 422.7 | 564.7 |

| T (K) | $\varepsilon_\infty$ | $\omega_{TO}$ (cm$^{-1}$) | $\Gamma_{TO}$ (cm$^{-1}$) | $\omega_{LO}$ (cm$^{-1}$) | $\Gamma_{LO}$ (cm$^{-1}$) |
|---|---|---|---|---|---|
| ~1630 | 1.34 | 163.8 | 74.8 | 191.8 | 98.8 |
| | | 377.4 | 139.0 | 453.8 | 160.7 |
| | | 624.4 | 350.2 | 650.3 | 187.0 |
| | | *5386.0* | *9507.8* | *8266.4* | *5542.2* |
| | | *10786.2* | *902.8* | *11825.7* | *205.8* |
| | | | | $\omega_{pl}$ (cm$^{-1}$) | |
| | | | 1225.2 | 755.3 | 1520.0 |

| T (K) | $\varepsilon_\infty$ | $\omega_{TO}$ (cm$^{-1}$) | $\Gamma_{TO}$ (cm$^{-1}$) | $\omega_{LO}$ (cm$^{-1}$) | $\Gamma_{LO}$ (cm$^{-1}$) |
|---|---|---|---|---|---|
| ~1670 | 1.2 | 172.4 | 71.3 | 193.1 | 93.2 |
| | | 366.2 | 110.1 | 390.6 | 209.4 |
| | | 575.9 | 592.5 | 643.3 | 307.3 |
| | | *3493.2* | *5636.7* | *8380.7* | *2137.0* |
| | | *11228.0* | *1582.7* | *13599.1* | *849.6* |
| | | | | $\omega_{pl}$ (cm$^{-1}$) | |
| | | | 8704.4 | 3165.3 | 9954.9 |

# Table IV

Parameters of the small polaron theory fits to the optical conductivity ErMnO$_3$ in the orthorhombic and cubic phase at 887 K, 1232 K, and ~1670 K (see text). Note that vibrational frequencies $\varpi_{phj}$ (j=1,2) are in agreement with the experimental (in brackets). These are $\omega_{TO}$ and $\omega_{LO}$ reststrahlen midvalue frequencies in Table III for the antisymmetric and symmetric (breathing) modes at 500-400 cm$^{-1}$ and ~600 cm$^{-1}$ respectively. Electron-phonon interactions associated with the octahedral symmetric and antisymmetric modes are enough to account for the measured optical conductivity. The DC conductivities, $\sigma_{DC}$, were kept fixed.

| T K | $\sigma_{DC}$ (ohm$^{-1}$-cm$^{-1}$) | $\eta_1$ | $\varpi_{ph1}$ (cm$^{-1}$) | $\eta_2$ | $\varpi_{ph2}$ (cm$^{-1}$) |
|---|---|---|---|---|---|
| **1255** | 123.5 | 15.1 | 630.1 (630.1) | 6.7 | 503.2 (480.8) |
| **882** | 368.3 | 10.5 | 622.4 (639.4) | 9.1 | 459.5 (483.4) |
| **~1670** | 4875 | 9.4 | 604.1 | 10.1 | 383.3 |

# Table V

Parameters used in the bipolaron simulation for the optical conductivity of cubic ErMnO$_3$ defined by eq. (13). Note that we have used the longitudinal optical frequencies of the octahedral breathing mode at ~640 cm$^{-1}$ (Table III) as fixed parameters in the calculation of bipolaron optical conductivities.

| T (K) | $\varepsilon_\infty$ | 2*E$_{bipolaron}$ (cm$^{-1}$) | $\sigma_{DC}$ ($\Omega^{-1}$cm$^{-1}$) | $\omega_{LO}$ (vib) (cm$^{-1}$) |
|---|---|---|---|---|
| 1384 | 1.05 | 5425 | 415 | 639.2 |
| 1425 | 1.02 | 5475 | 355 | 637.6 |
| 1523 | 1.03 | 6497 | 241 | 660.5 |
| 1563 | 1.14 | 6560 | 140 | 640.1 |
| ~1630 | 1.34 | 5690 | 238 | 650.3 |
| ~1670 | 1.20 | 4380 | 3070 | 643.3 |

# Figure Captions

**Figure 1** (color online) X-ray (CuKα) diffraction pattern for room temperature O-ErMnO$_3$. The inset corresponds to a view of the orthorhombic perovskite structure.

**Figure 2** (color online) Overall temperature dependence of O-ErMnO$_3$ infrared spectra: a) near normal reflectivity at 12 K; experimental: dots, full line: fit. Inset (a): Second order feature likely coupled to the Jahn-Teller distortion merging into the high temperature background at near the orbital disordered transition temperature. Inset (b): Temperature dependent lowest frequency band tail originated in inhomogeneities, experimental: dots, full line: Gaussian fit. The Er$^{3+}$ Kramer´s doublet transitions is shown as a secondary structure at slight higher frequencies (see text and table I). For better viewing the spectra have been vertically offset. b) High temperature O-ErMnO$_3$ near normal reflectivity (full line) and 1-emissivity (circles) in the orthorhombic phase denoting a gradual increase of orbital disorder rendering the "O" phase net higher temperature-higher lattice symmetry. Note that laser heated samples at ambient temperature are too noisy being too cold for emitting efficiently but starting at ~450 K both, reflection and emission, are in excellent agreement. For better viewing the spectra have been vertically shifted relative to each other.

**Fig. 3** (color in line) a) As measured temperature dependent paramagnon (short range antiferromagnetic correlated) zone center spin resonance. Inset: Several order higher intensity Er$^{3+}$ crystal field contribution. Note that the sharpening in the band profile at the lowest temperature may signal the onset of the Rare Earth ordering. b) Field dependent profiles at 5 K with the interference pattern extracted.

**Fig. 4** (color online) a) Far infrared temperature dependent phonon profiles in the orthorhombic and cubic phase. Inset: Same scale profiles at the transition temperature, b) corresponding mid-infrared spectra showing the emerging bipolaron band at the cubic phase transition (see text). Inset: complete spectra showing higher frequency interband transitions.

**Figure 5** (color online) **a)** Semilog plot for orthorhombic O-ErMnO$_3$ far infrared 1-emissivity at 882 K. Note that the gradual increase of orbital disorder toward the "O" phase shown in b) (1225 K) only results in a relative limited profile smoothing and phonon softening. The two arrows point to the merge of two lattice phonons at the phase transition (see also Fig. 4) (dots: experiment, full line: fit) Inset: full 1-emissivity infrared spectra relative flat response. Lower panels: real part optical conductivity at 882 K and 1125 K (full line: experimental, open circles: Fit after eq. (9)). The fit assumes that conductivity is the sum of two Gaussian-like (eq. 9) contributions (drawn in full lines); each calculated at a phonon frequency $\varpi_j$ and at a temperature T (see text and table IV). Inset: Same fits against the complete experimental optical conductivity.

**Figure 6** (color online) **a)** Semilog plot for cubic ErMnO$_3$ far infrared 1-emissivity at 1384 K and b) 1425 K in the cubic phase where the infrared active phonons are also depicted (after Ref. 19). Two arrows point to the merge of lattice phonons (see also Fig. 3). The circle at ~600 cm$^{-1}$ points to the anomalous weaker band profile. Inset: full 1-emissivity infrared spectra showing the distinctive mid infrared bipolaron profile in linear scale (dots: experiment, full line: fit). Lower panels: real part optical conductivity at 1384 K and 1425 K. (full line: experimental, open circles: fits after eq (13)). Inset: Same fits against the complete experimental optical conductivity.

**Figure 7** (color online) **a)** Semilog plot for cubic ErMnO$_3$ far infrared 1-emissivity at 1523 K and b) 1563 K in the cubic phase showing the emerging overdamped Drude term (square) and the screening increase in the ~600 cm$^{-1}$ phonon (dots: experiment, full line: fit). Inset: complete 1-emissivity infrared spectra showing the distinctive bipolaron profile in linear scale. Lower panels: real part optical conductivity at 1523 K and 1563 K (full line: experimental, open circles: fits after eq (13)). Inset: Same fits against the complete experimental optical conductivity.

**Figure 8** (color online) **a)** Semilog plot for cubic $ErMnO_3$ far infrared 1-emissivity at ~1630 K and b) ~1670 K showing the thermal driven insulator-metal transition (dots: experiment, full line: fit, squares: outlined Drude term). Note the now total screening of the ~600 $cm^{-1}$ phonon at ~1670 K. Inset: complete 1-emissivity infrared spectra. Lower panels: real part optical conductivity at ~1630 K (full line: experimental, open circles: fits after eq (9)) At ~1670 K it is shown the plain superposition of the two approaches suggesting coexistence (full line: experimental, open circles: fit after eq (8)), solid triangles: fit after eq. (9) reproducing the frequency downshifted bipolaron). Inset: (left panel) same fits against the complete experimental optical conductivity; (right panel) bipolaron temperature dependent peak position after Table V.

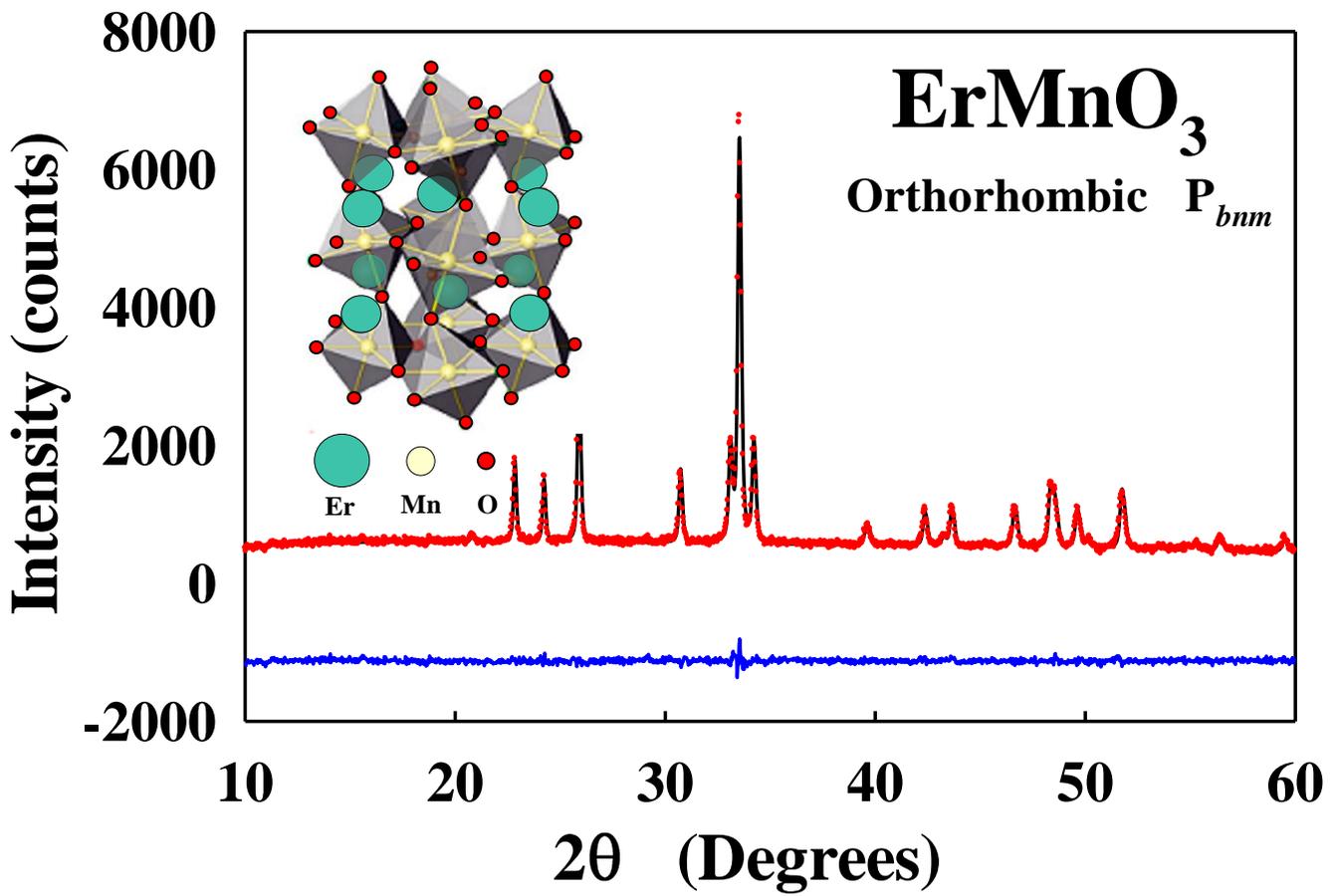

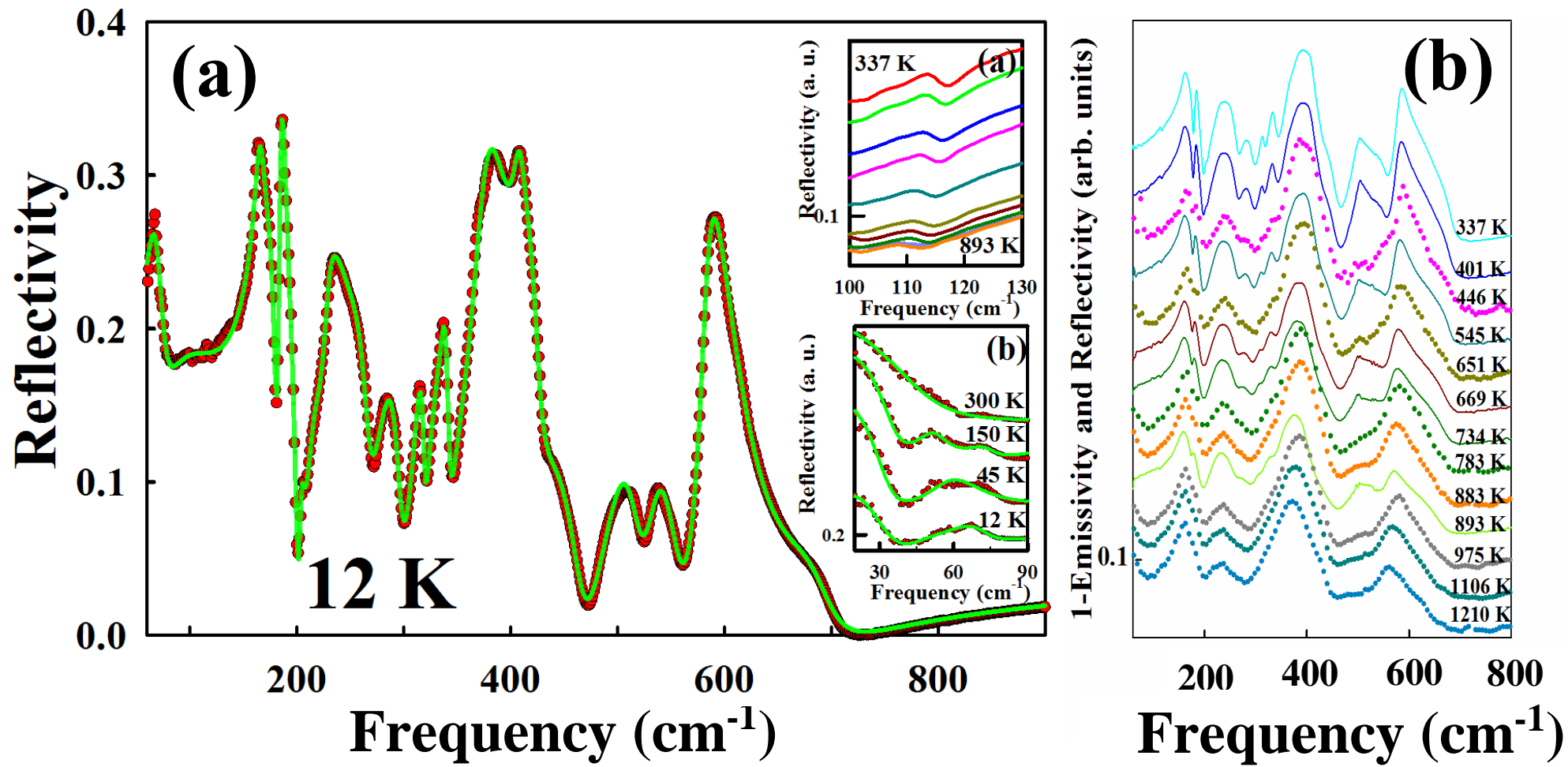

Massa et al
Fig. 2

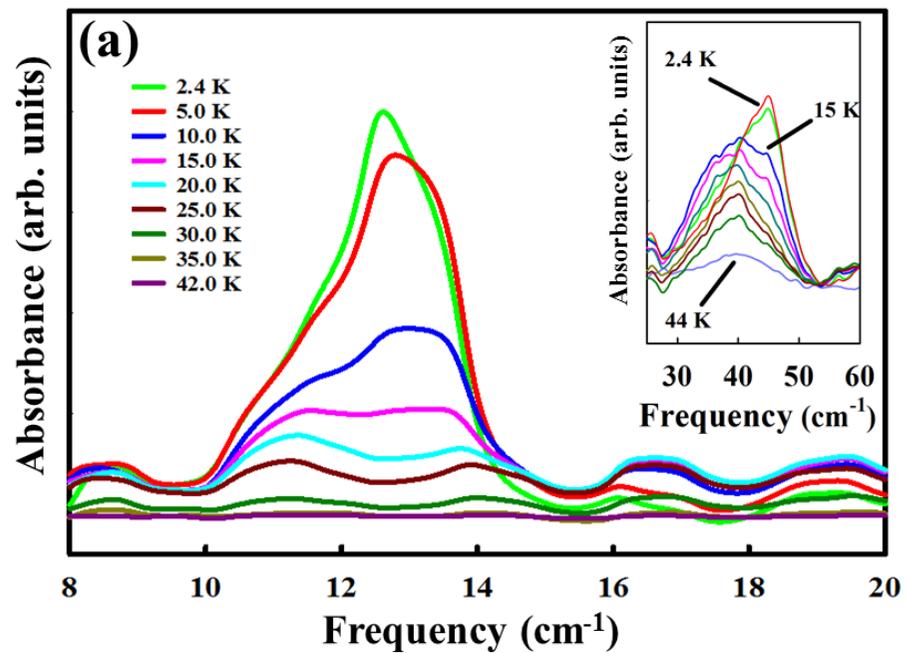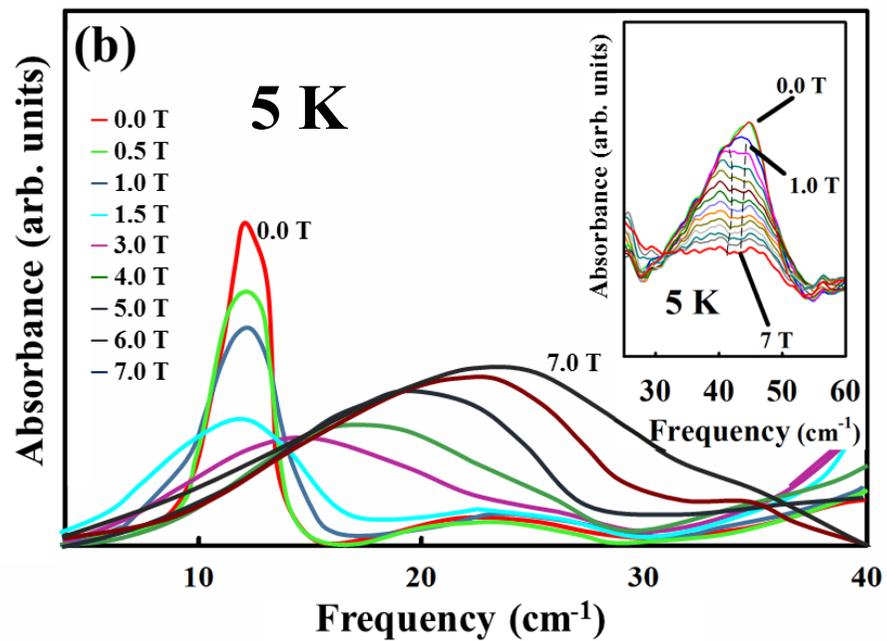

Massa et al
Fig. 3

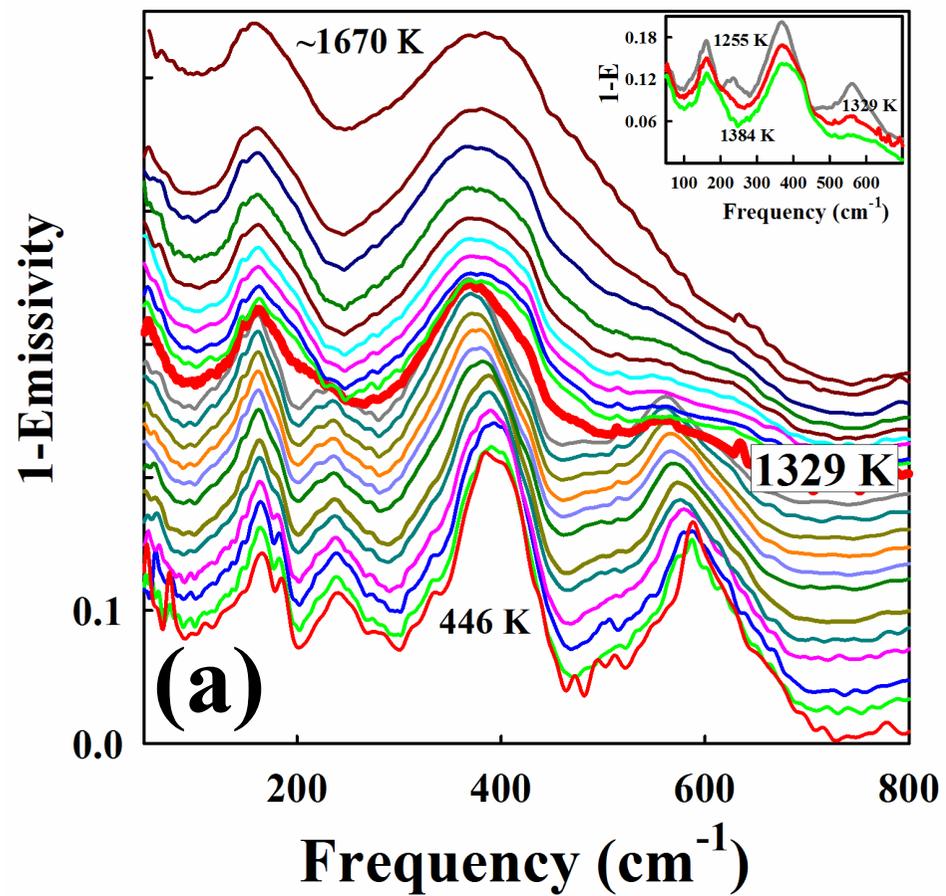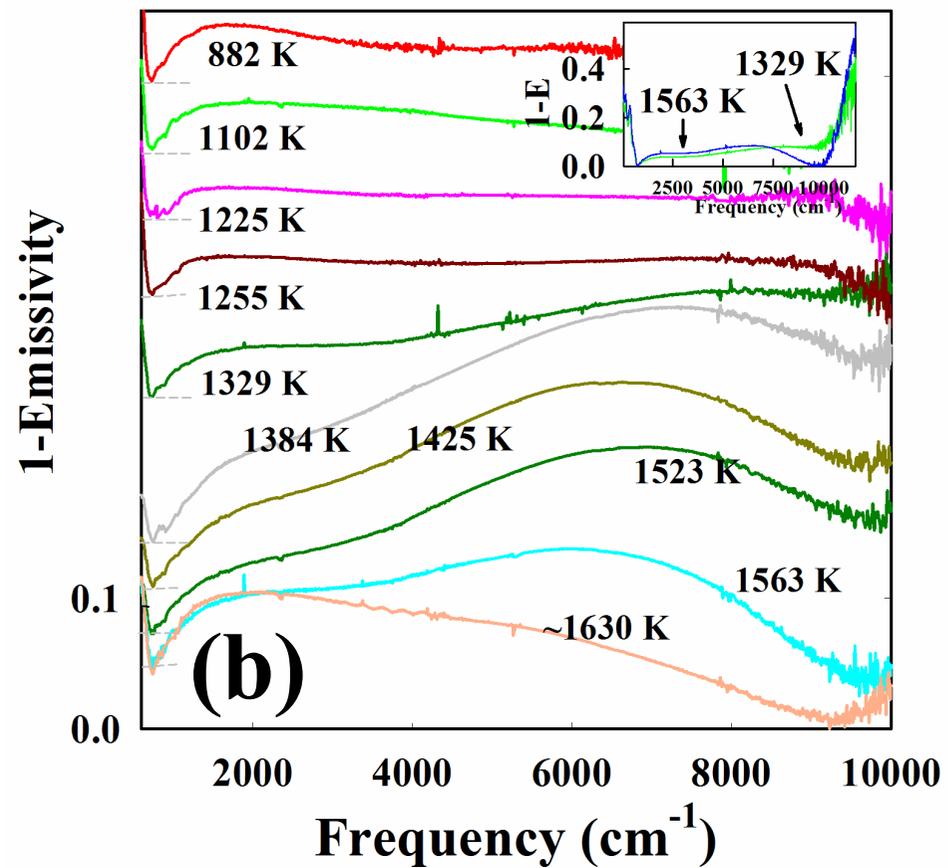

Massa et al
Fig. 4

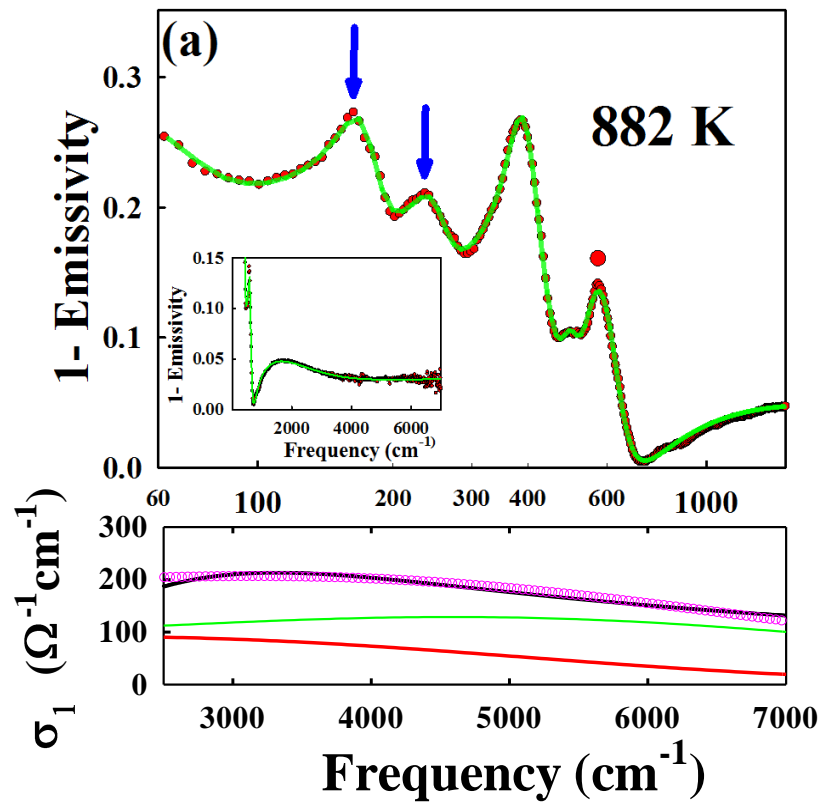
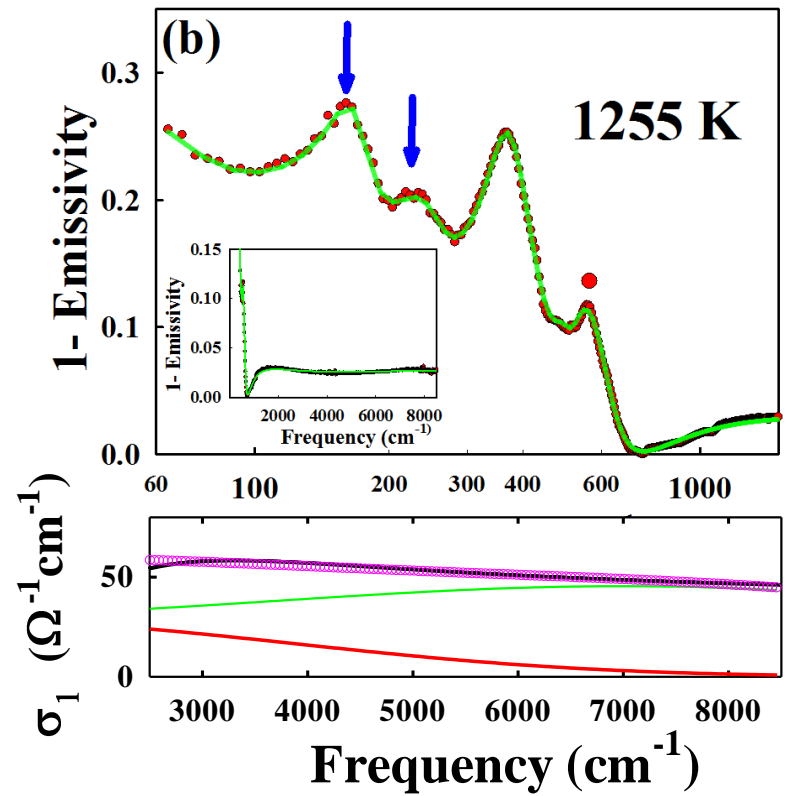

Massa et al
Fig. 5

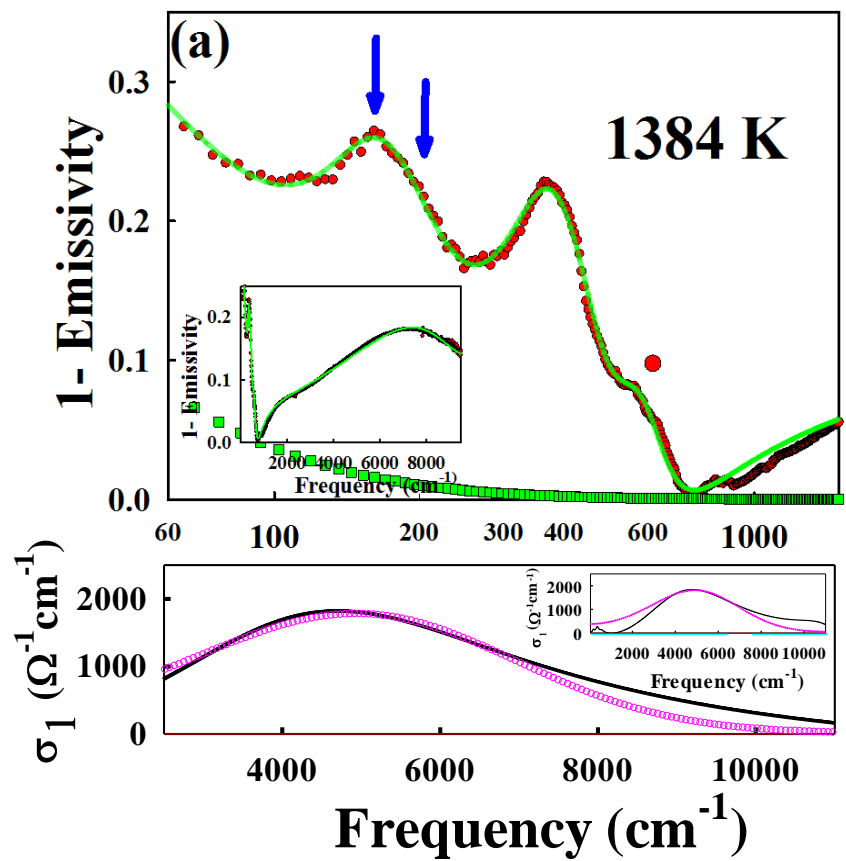
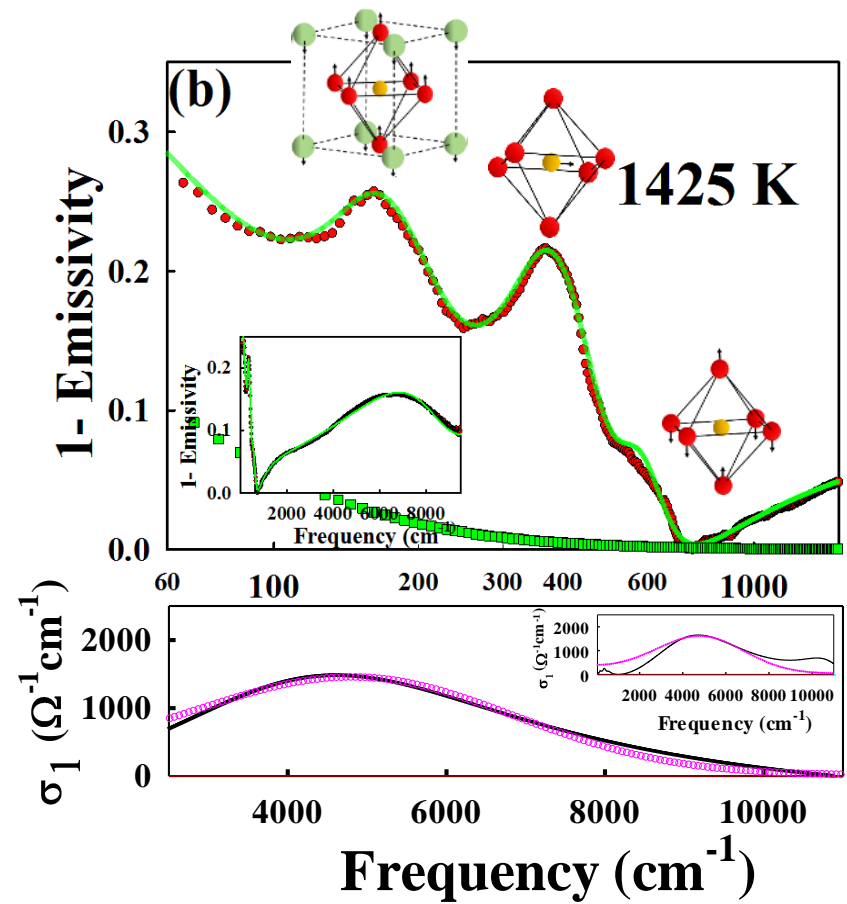

Massa et al
Fig. 6

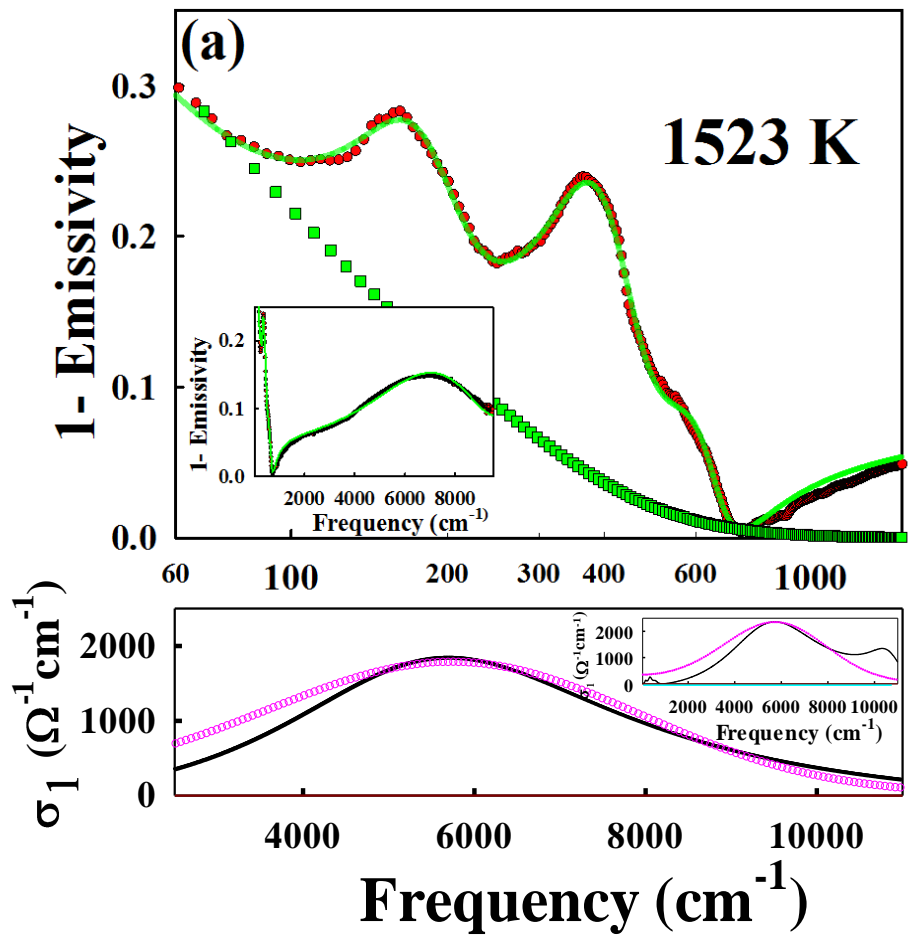
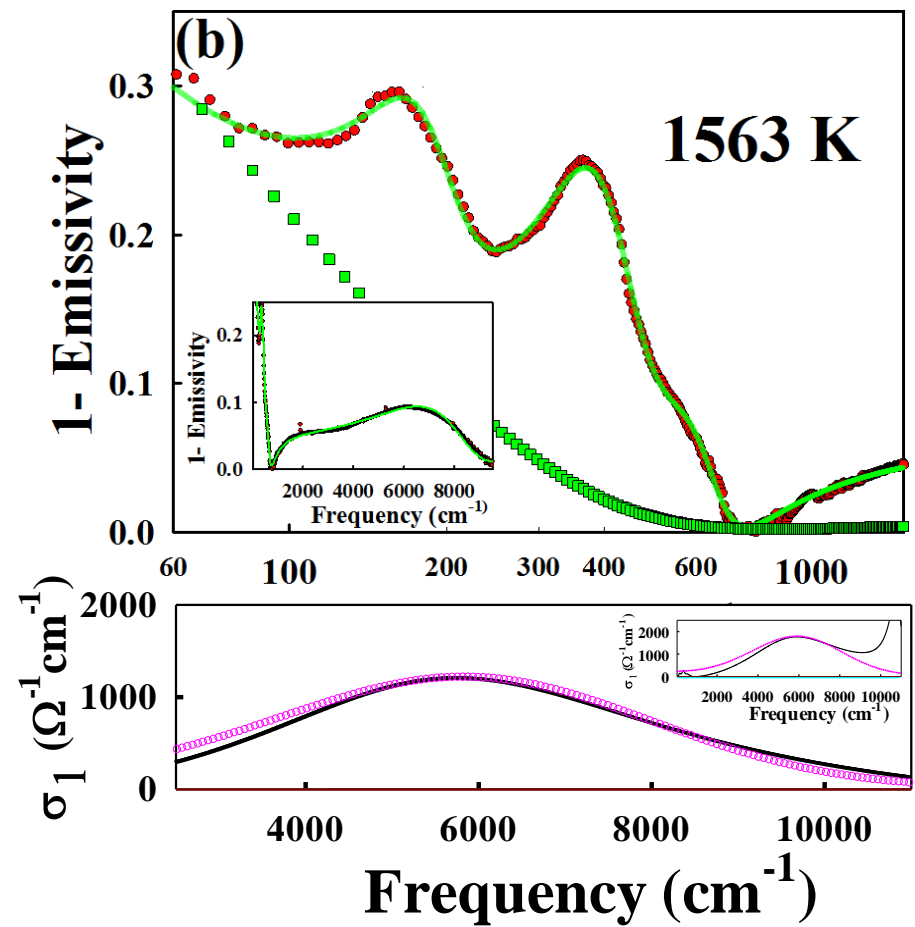

Massa et al
Fig. 7

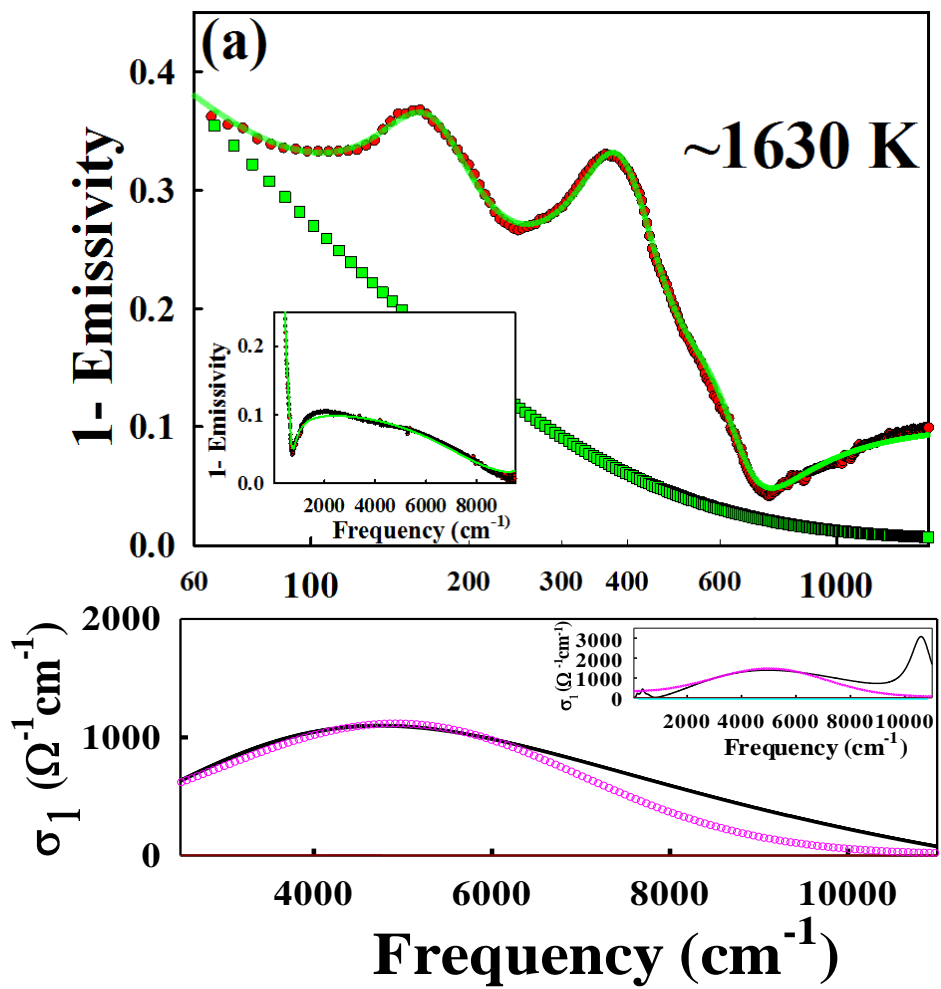
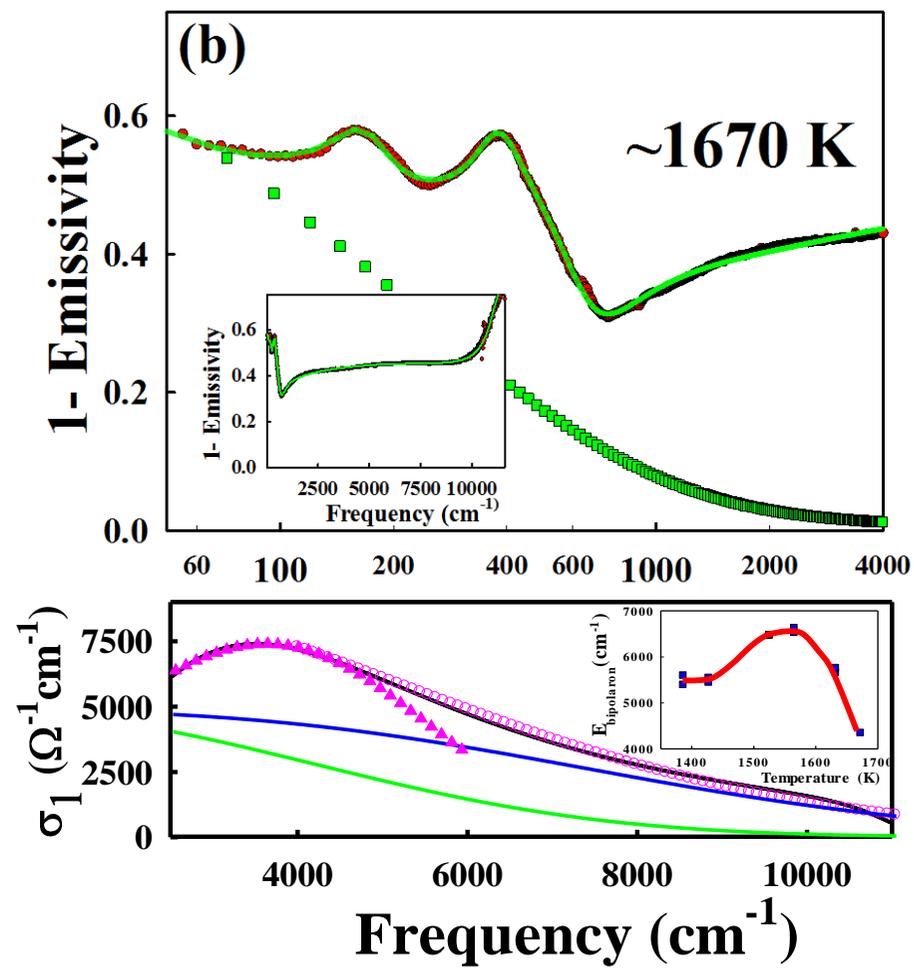

Massa et al
Fig. 8

47